\documentclass[preprint,a4,epsf]{revtex4}
\usepackage{graphicx,graphics} % Include figure files

\def\s{\sigma}

\def\up{\uparrow}
\def\dd{\downarrow}

\begin{document}
\title{Quantum theory of tunneling magnetoresistance in GaMnAs/GaAs/GaMnAs
heterostructures}
\author{Alireza Saffarzadeh}
\altaffiliation[Corresponding author. ]{E-mail:
a-saffar@tehran.pnu.ac.ir}
\author{Ali A. Shokri}
\address{Department of Physics, Tehran
Payame Noor University, Fallahpour St., Nejatollahi St., Tehran,
Iran}
\date{\today}

\begin{abstract}
Using a quantum theory including spin-splitting effect in diluted
magnetic semiconductors, we study the dependence of tunneling
magnetoresistance (TMR) on barrier thickness, temperature and
applied voltage in GaMnAs/GaAs/GaMnAs heterostructures. TMR ratios
more than 65\% are obtained at zero temperature, when one GaAs
monolayer ($\approx$ 0.565 nm) is used as a tunnel barrier. It is
also shown that the TMR ratio decreases rapidly with increasing
the barrier thickness and applied voltage, however at high
voltages and low thicknesses, the TMR first increases and then
decreases. Our model calculations well explain the main features
of the recent experimental observations.
\end{abstract}


\maketitle

\newpage
\section{Introduction}
In the past few years, tunneling magnetoresistance (TMR) in
magnetic tunnel junctions (MTJs) has attracted much attention due
to the possibility of its application in magnetic memories,
magnetic field sensors and quantum computing devices
\cite{Mood95,Daugh}. In a common MTJ, the TMR between two
ferromagnetic electrodes separated by an insulator layer depends
on the spin polarization of the ferromagnetic electrodes and spin
dependent tunneling rates \cite{Slon,Yu,Wang}. For such
structures, much efforts have been made to increase the TMR
ratio, which can be defined by the relative change in the current
densities. Recently, a large TMR ratio as much as 60\% has been
reported in Co-Fe/Al-O/Co-Fe junctions \cite{Tsunoda} which
opened a new era of technical applications of the tunnel
junctions.

Using ferromagnetic semiconductors (FMSs), such as EuS, the TMR
has also been investigated in single
\cite{Lec,Saffar1,Shokri1,Shokri2} and double
\cite{Worl,Filip,Saffar2,Wilc} magnetic barrier junctions. In
these structures, the FMSs which act as spin filters are used as
tunnel barriers. Therefore, the MTJs using these materials, are
able to produce highly spin-polarized current and high values for
TMR. However, the Cure temperature in most of FMSs is much lower
than room temperature \cite{Maug}.

On the other hand, coexistence of ferromagnetism and
semiconducting properties in diluted magnetic semiconductors
(DMSs) has opened the prospect of developing devices which make
it possible to combine the information processing and data
storage in the same material \cite{Ohno98,Dietl}. In the magnetic
semiconductors, the exchange interaction between the itinerant
carriers in the semiconducting band and the localized spins on
magnetic impurity ions leads to spectacular magneto-optical
effects, such as giant Faraday rotation or Zeeman splitting. Among
different types of DMSs, Ga$_{1-x}$Mn$_x$As is one of the most
suitable materials for use in spintronic devices. In this kind of
materials, a fraction of the group-III sublattice is replaced at
random by Mn ions which act as acceptors. According to the strong
Hund coupling, the five $3d$ electrons in a Mn$^{2+}$ ion give
rise to a localized magnetic moment. The DMSs based on III-V
semiconductors, like Ga$_{1-x}$Mn$_x$As compounds, exhibit
ferromagnetism with Curie temperature as high as 110 K for Mn
doping of around $x$=0.05. The origin of ferromagnetism in the
GaAs-based DMSs can be demonstrated through the $p$-$d$ exchange
coupling between the itinerant holes in the valence band of GaAs
and the spin of magnetic impurities \cite{Ohno98,Dietl,Konig}. In
spite of low Curie temperature of Ga$_{1-x}$Mn$_x$As, the use of
Ga$_{1-x}$Mn$_x$As-based III-V heterostructures as MTJs presents
several advantages. The possible advantages include not only
simple integration with existing semiconductor technology, but
also relatively straightforward fabrication of high-quality
epitaxial structures, easily controlled physical and electronic
structure, and the integration of quantum heterostructures which
is easier than in any other materials system \cite{Hayashi1}.

The spontaneous magnetization and the feasibility of preparing
GaAs-based DMS \cite{Hayashi} are stimulated much attention for
studying the spin-polarized transport phenomena in
Ga$_{1-x}$Mn$_x$As/(GaAs or AlAs) heterostructures. Recently,
Chiba \emph{et al.} \cite{Chiba} observed a TMR ratio of 5.5$\%$
at 20 K in a GaMnAs tunnel junction. Also, Tanaka and Higo
\cite{Tanaka} reported TMR ratios more than 70\% at 8 K in
GaMnAs/AlAs/GaMnAs heterostructures. Moreover, recent analyses of
temperature dependence of current-voltage characteristics of
GaMnAs/GaAs/p-GaAs p-i-p diodes have shown that GaAs intermediary
layer which is a nonmagnetic semiconductor (NMS), acts as a
barrier (87-140 meV) for holes injected from GaMnAs
\cite{Ohno2002}. In this regard, a large TMR ratio of 290\% has
been observed at 0.39 K in GaMnAs/GaAs/GaMnAs heterostructures,
around zero applied bias \cite{Chiba04}. Such a high TMR ratio is
a result of high spin polarization of DMS electrodes and the high
quality of sample and interfaces between GaMnAs and GaAs layer.
Interlayer exchange interaction is another effect which has also
been studied in Ga$_{0.96}$Mn$_{0.04}$As/GaAs superlattices,
where an oscillatory behavior with varying thickness of GaAs
spacer is predicted \cite{Mathieu}.

From a theoretical point of view, the TMR has also been
investigated in GaMnAs/AlAs/GaMnAs heterostructures. Tanaka and
Higo \cite{Tanaka} calculated the dependence of TMR on AlAs
thickness using the tight-binding theory including that the
parallel wave vector of carriers is conserved in tunneling
process. The results showed that the TMR ratio decreases rapidly
with increasing barrier thickness. Also, Tao \emph{et al.}
\cite{Tao} studied the AlAs thickness dependence of TMR using the
transfer matrix approach. Their treatment, however is somewhat
questionable, because they did not apply the boundary conditions
for derivatives of wave functions correctly. When the materials
are different, it is the normal mass flux that must be continuous.

In the present work, we study theoretically the spin-polarized
transport in GaMnAs/GaAs/GaMnAs tunnel junctions by considering
the effect of spontaneous magnetization of GaMnAs layers. The
temperature dependence of the spin splitting energy is calculated
for DMSs within the mean-field approximation. Using the transfer
matrix method and the nearly-free-carrier approximation we will
study the TMR and spin polarization of tunneling carriers in
terms of the barrier thickness and bias voltage at all
temperatures. We assume that the carrier wave vector parallel to
the interfaces and the carrier spin are conserved in the
tunneling process through the whole system.

The paper is organized as follows. In section 2, the model is
described and then the spin polarization and the TMR for
DMS/NMS/DMS structures are formulated. Next, the temperature
dependence of spin splitting energy is obtained for DMS, within
the mean-field framework. In section 3, numerical results
obtained for a typical tunnel junction are discussed. The results
of this work are summarized in section 4.

\section{Description of model and formalism}
In order to investigate the spin-dependent transport properties in
MTJs based on DMS materials, we consider two semi-infinite DMS
electrodes separated by a NMS layer, which acts as a barrier, in
the presence of DC applied bias $V_a$. For simplicity, we assume
that the two DMS electrodes are made of the same materials and
all of the interfaces are flat. In the framework of the
parabolic-band effective mass approximation, the longitudinal part
of the one-hole Hamiltonian can be written as
\begin{equation}\label{H}
H_x=-\frac{\hbar^2}{2m_j^*}\frac{d^2}{dx^2}+U_j(x)-{\bf h}_j^{\rm
MF}\cdot{\bf\s} ,
\end{equation}
where $m^*_j$ ($j$=1-3) is the hole effective mass in the $j$th
layer, and
\begin{equation}\label{U}
U_j(x)=\left\{\begin{array}{cc}
0, & x<0 \ ,\\
E_{\rm F}+\phi-eV_ax/d, & 0<x<d\  ,\\
-eV_a, & x>d\  ,\\
\end{array}\right.
\end{equation}
where $d$ is the barrier thickness, $E_{\rm F}$ is the Fermi
energy in the left electrode, measured from the middle point
between the edges of the two spin subbands, and $\phi$ is the
barrier height measured from the Fermi level. $-{\bf h}_j^{\rm
MF}\cdot{\bf\s}$ is the internal exchange energy where ${\bf
h}_j^{\rm MF}$ is the molecular field in the $j$th DMS electrode
and ${\bf\s}$ is the conventional Pauli spin operator.

The Schr\"odinger equation for a biased barrier layer can be
simplified by a coordinate transformation whose solution is the
linear combination of the Airy function Ai[$\rho(x)$] and its
complement Bi[$\rho(x)$] \cite{Abram}. Considering all three
regions of the DMS/NMS/DMS junction, the eigenfunctions of the
Hamiltonian (\ref{H}) with eigenvalue $E_x$ have the following
forms:
\begin{equation}\label{psi}
\psi_{j,\s}(x)=\left\{\begin{array}{cc}
A_{1\s}e^{ik_{1\s}x}+B_{1\s}e^{-ik_{1\s}x}, & x<0 \ ,\\
A_{2\s}{\rm Ai}[\rho(x)]+B_{2\s}{\rm Bi}[\rho(x)], &
0<x<d ,\\
A_{3\s}e^{ik_{3\s}x}+B_{3\s}e^{-ik_{3\s}x},& x>d ,\\
\end{array}\right.
\end{equation}
where,
\begin{equation}
k_{1\s}=\sqrt{2m_1^*(E_x+h_{0}\s)}/\hbar\ ,
\end{equation}
\begin{equation}
k_{3\s}^{(\Gamma)}=\sqrt{2m_3^*(E_x+eV_a+\Gamma h_{0}\s)}/\hbar\ ,
\end{equation}
are the hole wave vectors along the $x$ axis. Here, $\s$ are the
hole spin components $\pm 1$ (or $\up,\dd$), $h_{0}=|{\bf
h}_j^{\rm MF}|$ and $\Gamma=+1 (-1)$ for parallel (antiparallel)
alignment of the magnetizations. The coefficients $A_{j\s}$ and
$B_{j\s}$ are constants to be determined from the boundary
conditions, while
\begin{equation}\label{rho}
\rho(x)=-\frac{d}{eV_a\lambda}(E_{\rm
F}+\phi-E_x-\frac{x}{d}eV_a) \ ,
\end{equation}
with
\begin{equation}
\lambda=\left[\frac{-\hbar^2d}{2m^*_2eV_a}\right]^{1/3} \ .
\end{equation}

Although the transverse momentum ${\bf k}_\parallel$ is omitted
from the above notations, the summation over ${\bf k}_\parallel$
is carried out in our calculations.

Upon applying the boundary conditions such that the wave functions
and their first derivatives are matched at each interface point
$x_j$ , i.e.  $\psi_{j,\s}(x_j)=\psi_{j+1,\s}(x_j)$ and
$(m^*_j)^{-1}[d\psi_{j,\s}(x_j)/dx]=(m^*_{j+1})^{-1}[d\psi_{j+1,\s}(x_j)/dx]$,
we obtain a matrix formula that connects the coefficients
$A_{1\s}$ and $B_{1\s}$ with the coefficients $A_{3\s}$ and
$B_{3\s}$ \cite{Saffar1}. Since there is no reflection in region
3, the coefficient $B_{3\s}$ in Eq. (\ref{psi}) is zero and the
transmission coefficient of the spin $\s$ hole, which is defined
as the ratio of the transmitted flux to the incident flux can be
written as
\begin{equation}\label{T}
T_\s^{(\Gamma)}(E_x,V_a)=\frac{4m_1^{*}m_3^{*}k_{1\s}k_{3\s}^{(\Gamma)}}{(\pi
\lambda m_2^*)^2} \left[(\alpha k_{1\s}k_{3\s}^{(\Gamma)}+\delta
m_1^{*}m_3^{*})^2+(\beta m_3^{*}k_{1\s}-\gamma
m_1^{*}k_{3\s}^{(\Gamma)})^2\right]^{-1} \ ,
\end{equation}
where the following abbreviations are used
\begin{eqnarray}
\alpha={\rm Ai}[\rho(0)]{\rm Bi}[\rho(d)]-{\rm Bi}[\rho(0)]{\rm Ai}[\rho(d)] ,\\
\beta=\frac{1}{\lambda m_2^*}
\{{\rm Ai}[\rho(0)]{\rm Bi}'[\rho(d)]-{\rm Bi}[\rho(0)]{\rm Ai}'[\rho(d)]\} ,\\
\gamma=\frac{1}{\lambda m_2^*}
\{ {\rm Ai}'[\rho(0)]{\rm Bi}[\rho(d)]-{\rm Bi}'[\rho(0)]{\rm Ai}[\rho(d)]\} ,\\
\delta=\frac{1}{(\lambda m_2^*)^2}\{{\rm Ai}'[\rho(0)]{\rm
Bi}'[\rho(d)]-{\rm Bi}'[\rho(0)]{\rm Ai}'[\rho(d)]\} \ .
\end{eqnarray}
Here, ${\rm Ai}'[\rho(x)]$ and ${\rm Bi}'[\rho(x)]$ are the first
derivatives of the Airy functions.

\subsection{Current densities and TMR}
The spin-dependent current density for a MTJ with a given applied
bias $V_a$ and at temperature $T$ can be calculated within the
nearly-free-hole approximation \cite{Duke}:
\begin{equation}
J_\s^{(\Gamma)}(V_a)=\frac{em^*_1k_BT}{4\pi^2\hbar^3}\int_{E_0^\s}^{\infty}
T_\s^{(\Gamma)}(E_x,V_a){\rm ln}\left\{\frac{1+{\rm
exp}\left[(E_{\rm F}-E_x)/k_BT \right]}{1+{\rm exp}\left[(E_{\rm
F}-E_x-eV_a)/k_BT \right]}\right\}dE_x \ ,
\end{equation}
and, at $T$=0 K,
\begin{equation}
J_\s^{(\Gamma)}(V_a)=\frac{em^*_1}{4\pi^2\hbar^3}\left[eV_a\int_{E_0^\s}^{E_{\rm
F}-eV_a} T_\s^{(\Gamma)}(E_x,V_a)dE_x+\int_{E_{\rm
F}-eV_a}^{E_{\rm F}}(E_{\rm F}-E_x)
T_\s^{(\Gamma)}(E_x,V_a)dE_x\right]\ ,
\end{equation}
where $k_B$ is the Boltzmann constant and $E^\s_0$ is the lowest
possible energy that will allow transmission and is given by
$E^{\up}_0 =\max\{-h_0,-eV_a-\Gamma h_0\}$ for spin-up holes and
$E^{\dd}_{0} =h_0$ for spin-down ones. It is clear that the
tunnel current is modulated by the magnetic configurations of the
both DMS electrodes.

The degree of spin polarization for the tunnel current is defined
by $P=(J_\up-J_\dd)/(J_\up+J_\dd)$, where $J_\up$ $(J_\dd)$ is the
spin-up (spin-down) current density. For the present structure, we
obtain this quantity when the magnetizations of two DMS electrodes
are in parallel alignment. For studying the TMR, the spin currents
in the both parallel and antiparallel alignments are calculated.
In this case, the TMR can be defined as
\begin{equation}\label{tmr}
\mbox{TMR}=\frac{(J_{\up}^{\rm p}+J_{\dd}^{\rm p})-(J_{\up}^{\rm
ap} +J_{\dd}^{\rm ap})}{J_{\up}^{\rm ap}+J_{\dd}^{\rm ap}} \ ,
\end{equation}
where $J_{\up,\dd}^{\rm p(ap)}$ corresponds to the current
density in the parallel (antiparallel) alignment of the
magnetizations in the FM electrodes, for a spin-up ($\up$) or
spin-down ($\dd$) hole.

\subsection{Temperature dependence of spin-splitting energy}
In the following, we develop the formalism for calculating of
temperature dependence of the spin splitting energy in the DMS
electrodes using the mean-field theory. At the first step, we
consider the magnetic Hamiltonian for one itinerant hole with spin
${\bf s}$ located at ${\bf r}$ and one Mn ion with spin ${\bf S}$
located at ${\bf R}$. Both of these magnetic subsystems can be
described as \cite{Sarma}
\begin{equation}
H^{\rm h}=J_{\rm pd}\sum _{I}{\bf s}\cdot{\bf S}_I\,\delta({\bf
r}-{\bf R}_I) \ ,
\end{equation}
and
\begin{equation}
H^{\rm Mn}=J_{\rm pd}\sum _{i}{\bf s}_i\cdot{\bf S}\,\delta({\bf
r}_i-{\bf R}) \ .
\end{equation}
Here, $J_{\rm pd}$ is the hole-Mn exchange coupling constant,
$\mu_B$ is the Bohr magneton, $g_{\rm h}$ and $g_{\rm Mn}$ are the
Land\'{e} g-factor for holes and Mn ions, respectively. In the
mean-field approximation, $H^{\rm h} \longrightarrow H^{\rm
h}_{\rm MF}\equiv g_{\rm h}\mu_B{\bf s}\cdot{\rm {\bf h}}^{\rm h}$
and $H^{\rm Mn} \longrightarrow H^{\rm Mn}_{\rm MF}\equiv g_{\rm
Mn}\mu_B{\bf S} \cdot {\rm {\bf h}}^{\rm Mn}$, where ${\bf h}^{\rm
h}$ and ${\bf h}^{\rm Mn}$ are respectively the effective magnetic
fields acting upon holes and magnetic impurities, and can be
given as
\begin{equation}
h^{\rm h}=\frac{1}{g_{\rm h}\mu_B}J_{\rm pd} N_{\rm Mn}\langle
S_z\rangle \ ,
\end{equation}
and
\begin{equation}
h^{\rm Mn}=\frac{1}{g_{\rm Mn}\mu_B}J_{\rm pd}{\rm M^h} \ .
\end{equation}
Here, $N_{\rm Mn}=4x/a_0^3$ is the density of Mn ions, with
$a_0^3$ being the unit cell volume. ${\bf M}^{\rm h}=\sum
_{i}\langle{\bf s}_i\rangle\,\delta({\bf r}_i-{\bf R})$ is the
magnetization density of the hole subsystem, which is assumed to
be uniform within the length scale of the magnetic interactions,
so the magnetic response of the the Fermi sea holes to the
effective field $h^{\rm h}$, is given by
\begin{equation}
{\rm M^h}=g_{\rm h}\mu_B s^2D^{\rm h}(E_{\rm F}) h^{\rm h} ,
\end{equation}
where the density of states of the hole gas with effective mass
$m^*$ and the hole concentration $p$ is $D^{\rm h}(E_{\rm
F})=(3\pi^2)^{-2/3}(3m^*/\hbar^2)p^{1/3}$. On the other hand, the
magnetic response of the impurity spin to the effective field
$h^{\rm Mn}$, is given by
\begin{equation}
{\rm M^{Mn}}=g_{\rm Mn}\mu_B N_{\rm Mn}\langle S_z\rangle= g_{\rm
Mn}\mu_B N_{\rm Mn}S\mathcal{B}_S\left(\frac{{g_{\rm Mn}\mu_BS
h^{\rm Mn}}}{k_BT}\right) ,
\end{equation}
where
\begin{equation}
\mathcal{B}_S(x)=(\frac{2S+1}{2S})\coth(\frac{2S+1}{2S}x)-
(\frac{1}{2S})\coth(\frac{1}{2S}x) \ ,
\end{equation}
is the Brillouin function. Therefore, within the spirit of a
mean-field framework the magnetization of Mn subsystem is then
given by
\begin{equation}
{\rm M^{Mn}}=g_{\rm Mn}\mu_BN_{\rm
Mn}S\mathcal{B}_S\left(\frac{J_{\rm pd}S}{2k_BT}{\rm M^h}\right) \
,
\end{equation}
which should be determined self-consistency with the hole
magnetization
\begin{equation}
{\rm M^h}=\frac{J_{\rm pd}D^{\rm h}(E_{\rm F})}{2g_{\rm
Mn}\mu_B}{\rm M^{Mn}} \ .
\end{equation}
Also, the ferromagnetic transition temperature $T_{\rm C}$ can be
obtained using the expansion for the Brillouin function
$\mathcal{B}_S(x)$, when $x\ll 1$. In this regard, $T_{\rm C}$ is
found as
\begin{equation}
T_{\rm C}=(J_{\rm pd}s)^2N_{\rm Mn}\frac{S(S+1)D(E_{\rm
F})}{3k_B}.
\end{equation}
Now, the temperature dependence of spin splitting energy in the
$j$th DMS electrode can be obtained as $\Delta_j=2|{\bf h}_j^{\rm
MF}| $, where, the molecular field is ${\bf h}_j^{\rm MF}=(J_{\rm
pd}/2g_{\rm Mn}\mu_B){\bf M}^{\rm Mn}_j$.

\section{Numerical results and discussions}
Numerical calculations have been carried out to investigate the
effects of barrier thickness, temperature and applied voltage on
spin-dependent transport in a typical
Ga$_{1-x}$Mn$_x$As/GaAs/Ga$_{1-x}$Mn$_x$As tunnel structure. We
have chosen Ga$_{1-x}$Mn$_x$As and GaAs because of the same
crystal structures and lattice constants \cite{Tanaka}. The
relevant parameters for Ga$_{1-x}$Mn$_x$As electrodes are chosen
as $a_0=5.65$ \AA, $E_{\rm F}$=0.2 eV \cite{Tanaka}, $J_{pd}$=0.15
eV-nm$^3$ \cite{Jungwirth}, $S$=5/2, $s$=1/2, $g_{\rm h}=g_{\rm
MN}$=2, $p=4.9\times 10^{20}\rm cm^{-3}$ \cite{Ohno99} and
$T_C=110$ K for a sample with $x=0.05$ \cite{Ohno98}. The suitable
parameter for the barrier height of GaAs is $\phi$=0.1 eV
\cite{Chiba04}.

It is important to note that in the valance band of
semiconductors, there are heavy hole and light hole bands which
are degenerate at the top of the valance band. Therefore, it
seems to be reasonable that we use of one value for the hole mass
at the structures with low Mn concentration. Hence, the effective
mass of all carriers are taken as $m^*$=0.16 $m_h$ ($m_h$ is the
free-hole mass), which is consistent with the experimental
parameters have been used in the present work.

In Fig. 1 we have presented the TMR as a function of the applied
voltage for several temperatures when $d=0.565$ nm. As it is
evident from the figure, at all temperatures the TMR decreases
monotonically as $V_a$ increases. Also, it is seen that, the bias
voltage values, where the TMR ratio reaches half its maximum,
varies from 55 to 90 mV depending on temperature.This value is
much smaller than that of metal-based single barrier MTJ
structures (300-400 mV) \cite{Mood95,Sousa,Lu}. At very low
voltages, for both parallel and antiparallel alignments, the
tunnel current densities vary linearly, therefore the TMR behaves
similar to a flat curve. However, with increasing the applied
voltage, due to the modification of the barrier shape, a nearly
parabolic dependence of currents on the voltage appears, and this
modifies the TMR curve. With increasing the temperature, the
spontaneous spin splitting in the valence band of the DMS layers,
decreases. In this case, the difference between current densities
in both parallel and antiparallel configurations is also reduced.
Thus, in such temperatures one can expect low values for TMR
ratio, as it is clear in the figure. The results show that the
discrepancy between the TMR(100 K) and TMR(0 K) decreases with
increasing the bias voltage. Therefore, the applied bias has a
strong influence on the TMR. The relative behavior of the curves
in Fig. 1, is qualitatively in good agreement with the recent
experimental results \cite{Chiba04}.

The barrier thickness dependence of TMR at different applied
biases has been shown in Fig. 2. For $V_a=$5 and 80 mV, with
increasing the barrier thickness, the TMR decreases rapidly from
its maximum value and always remains positive for various
thicknesses of the GaAs layer. However, for $V_a=$120 and 150 mV,
it is found that the TMR becomes negative in a certain range of
the barrier thickness, as it is shown in the inset. The origin of
this effect is related to the enhancement of the applied bias and
can be observed, when the applied bias is more than the barrier
height (100 meV). In fact, at high-voltage range ($eV_a>\phi$) a
quantum well will appear at GaAs/GaMnAs interface; thus, the
carriers at the Fermi level of the left electrode tunnel through
the GaAs barrier and the quantum well, into the right electrode.
In this regard, the carriers can cause standing waves in the
barrier, leading to a reduction and a sign reversal of the TMR,
as it is shown in the inset of Fig. 2. Also, with further
increasing of the applied voltage, the TMR exhibits a peak at low
thicknesses, which has been observed previously in
GaMnAs/AlAs/GaMnAs heterostructures \cite{Tanaka}.

The temperature dependence of the TMR and spin polarization is
also calculated in the present MTJ. The results are shown in Fig.
3 at $V_a$=5 mV when $d=0.565$ nm. From the figure, it is clear
that, both the TMR and spin polarization decrease with increasing
temperature, and vanish at 110 K, which is the Curie temperature
of the Ga$_{0.95}$Mn$_{0.05}$As layers. One can see that, the
highest value of TMR (about 70$\%$) and spin polarization (about
37$\%$) can be obtained at zero temperature. The inset shows the
normalized magnetization of Ga$_{0.95}$Mn$_{0.05}$As layers as a
function of temperature. The reduction of TMR and spin
polarization arises as a result of the dependence of spin
splitting energy on temperature. The comparison of the present
results and those obtained with experiment \cite{Chiba04} shows
somewhat different behavior at 10-30 K. Such discrepancy between
our theory and the experiment can be probably due to the effects
of magnetic anisotropy at the GaMnAs/GaAs interfaces and also
spin-flip scattering at low temperature in the GaMnAs electrodes.
The mean-field theory does not take into account the effects
mentioned here.

Our final comment addresses the role of Rashba spin-orbit
interaction in the present tunnel structure \cite{Mak1,Mak2}. The
Rashba effect produces a spin-flip scattering in the interfaces,
which affect on the transmission of each spin channel and can not
be solved separately. Therefore, a spin mixture occurs and the
currents become less spin polarized. However, the Rashba effect
is very small. In fact, with assuming the Rashba interaction in
the magnetic tunnel junctions, it has been shown that, the
spin-dependent transmission coefficients change very small in the
both Parallel and antiparallel configurations \cite{Mak2}. Hence,
the variation of spin polarization and TMR due to this effect is
nearly small. Therefore, one can omit this effect in such tunnel
heterostructure. That is, the Rashba effect does not have a
dominant influence on the spin polarization and TMR.

\section{Summary and conclusions}
On the basis of the effective mass approximation model including
the temperature dependence of spin splitting, we studied the
spin-polarized tunneling in
Ga$_{0.95}$Mn$_{0.05}$As/GaAs/Ga$_{0.95}$Mn$_{0.05}$As
heterostructure. The numerical results showed that, the TMR
strongly depends on GaAs barrier thickness, applied voltage and
temperature. At high voltages, with increasing the barrier
thickness, a peak and sign reversal observed in the TMR curves.
Also, we found that, both the TMR and spin polarization decrease
with increasing the temperature. The results obtained in this
work show the main features of the experiments and can be a base
of development for spin dependent tunneling electronic devices.

\newpage
\begin{figure}
\centering \resizebox{0.8\textwidth}{0.7\textheight}
{\includegraphics{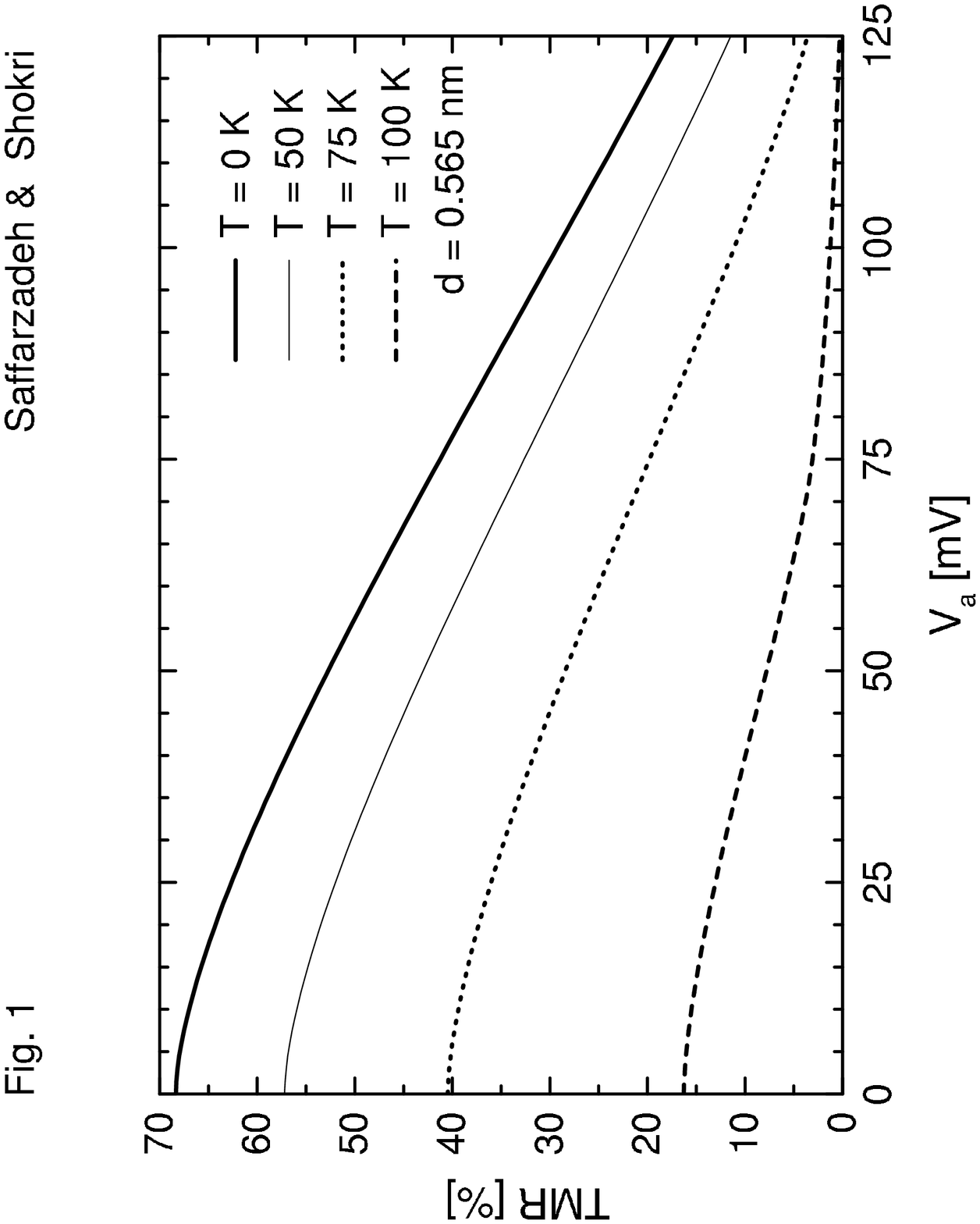}} \caption{Dependence of the TMR on the
applied voltage at different temperatures.}
\end{figure}
\newpage
\begin{figure}
\centering \resizebox{0.8\textwidth}{0.7\textheight}
{\includegraphics{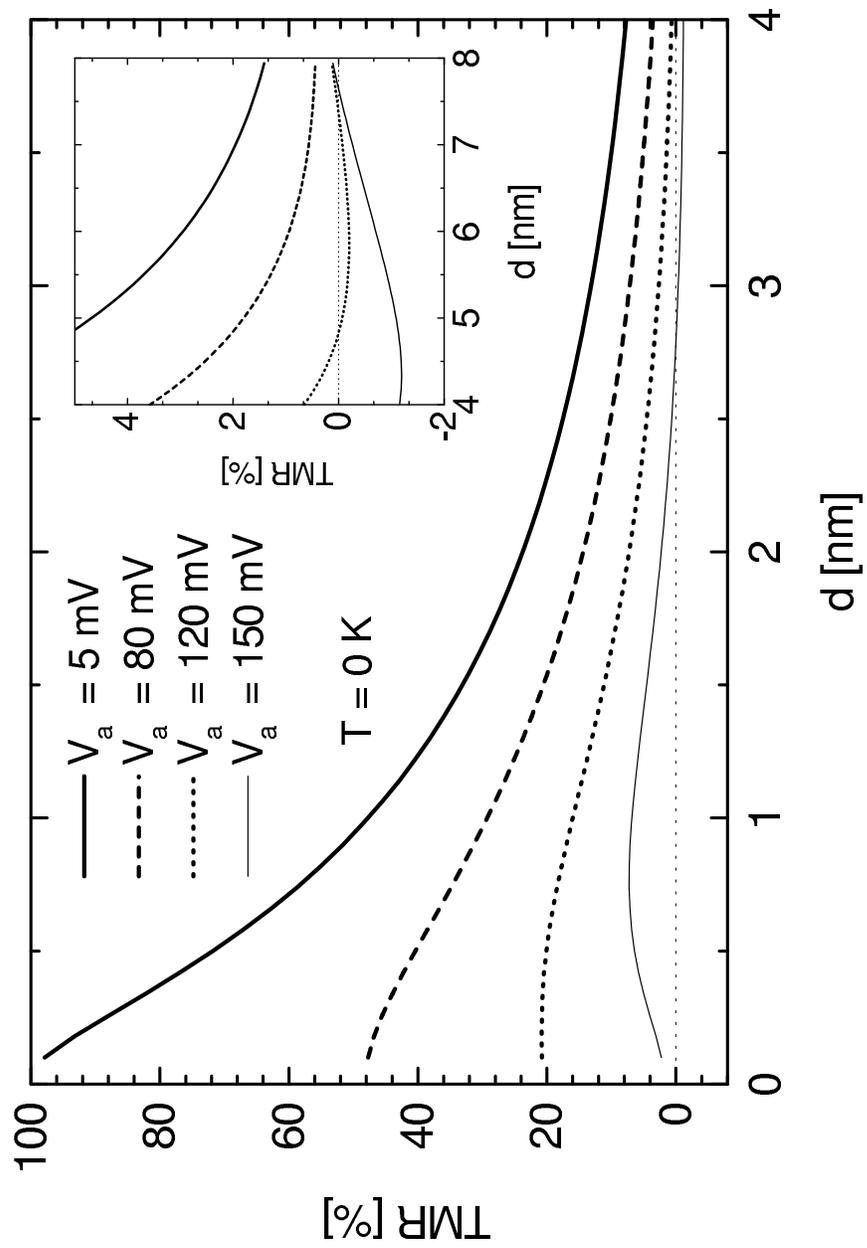}} \caption{Variations of the TMR as a
function of the barrier thickness at different temperatures.}
\end{figure}
\newpage
\begin{figure}
\centering \resizebox{0.8\textwidth}{0.7\textheight}
{\includegraphics{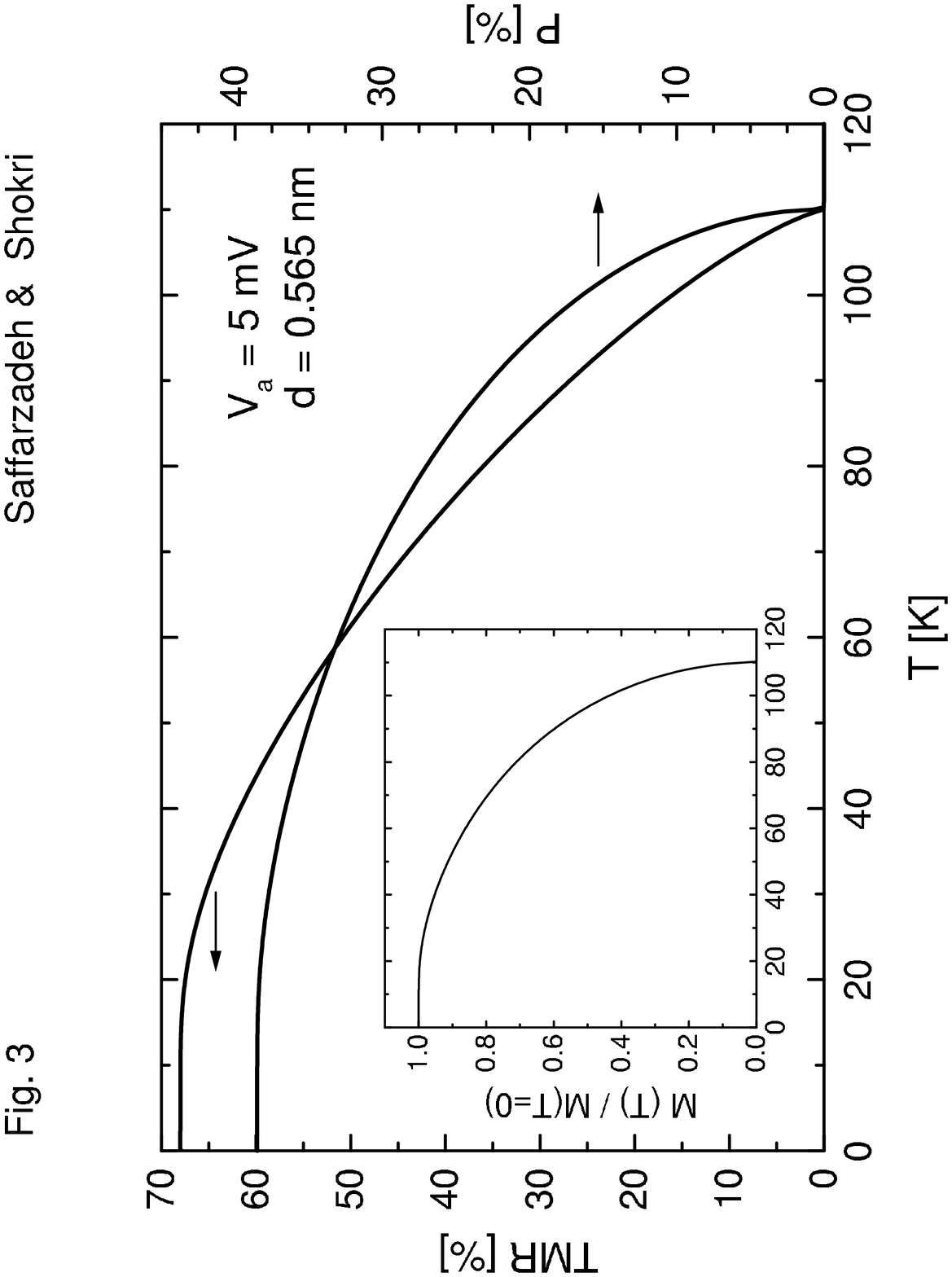}} \caption{Dependence of the TMR (left)
and spin polarization (right) as a function of temperature. The
inset shows temperature dependence of the normalized
magnetization of Ga$_{0.95}$Mn$_{0.05}$As electrodes.}
\end{figure}

\end{document}